\documentclass[structabstract]{aa}
\usepackage{graphicx}
\usepackage{lscape}
\def\approxgt{\mathrel{\hbox{\rlap{\lower.55ex \hbox {$\sim$}}
        \kern-.3em \raise.4ex \hbox{$>$}}}}
\def\approxlt{\mathrel{\hbox{\rlap{\lower.55ex \hbox {$\sim$}}
        \kern-.3em \raise.4ex \hbox{$<$}}}}
\begin{document}
   \title{The X-ray variability history of Markarian~3}

   \author{
          M.~Guainazzi
          \inst{1},
	  V.~La Parola
          \inst{2},
          G.~Miniutti
          \inst{3},
          A.~Segreto
          \inst{2},
          A.L.~Longinotti
          \inst{1}
          }

   \offprints{M.Guainazzi}

   \institute{
              $^1$European Space Astronomy Centre of ESA P.O.Box 78,
              Villanueva de la Ca\~nada, E-28691 Madrid, Spain \\
	      \email{Matteo.Guainazzi@sciops.esa.int} \\
              $^2$INAF – IASF Palermo, via Ugo La Malfa 153, 90146 Palermo, Italy \\
              $^2$Centro de Astrobiologia (CSIC-INTA), Dep. de Astrofisica; ESA, PO Box 78,
              E-28691, Villanueva de la Ca\~nada, Madrid, Spain)
              }

   \date{Received ; accepted }

   \abstract{The unified scenario for Active Galactic Nuclei (AGN)
            postulates that our orientation with respect to a parsec-scale azimuthally-symmetric gas and
            dust system causes their different
            phenomenology in the optical/UV and X-ray bands. Only
            recently, high-resolution radio (VLBI) and IR interferometric observations
            have provided direct constraints on size and structure of this obscuring
            system (known historically as the ``torus'').
            On the other hand, variability of optically-thick X-ray
            absorption and reprocessing in heavily obscured AGN
            often probe smaller scales, down to the Broad Line Region and beyond.}
	    {We aim at constraining the geometry of the reprocessing matter in the nearby
            prototypical Seyfert~2 Galaxy Markarian~3 by studying the time evolution of spectral
            components associated to the primary AGN emission and to its Compton-scattering.}
	    {We have analyzed archival spectroscopic observations of Markarian~3 taken over the last
            $\simeq$12~years with the XMM-Newton, Suzaku and Swift observatories, as well as
            data taken during a monitoring campaign activated by us in 2012.}
	    {The timescale of the Compton-reflection component variability (originally
            discovered by ASCA in the mid-'90s) is $\approxlt$64~days. This upper limit
            improves by more than a factor of 15 previous estimates
            of the Compton-reflection variability timescale for this source.
            When the light curve of
            the Compton-reflection continuum in the 4-5~keV band is correlated with the 14--195~keV
            Swift/BAT curve a delay $\approxgt$1200~days is found.
            These results are dependent on the model used to fit
            the spectra, although the detection of the Compton-reflection component variability
            is independent of the range of models employed to fit the data.
            Reanalysis of an archival Chandra
            image of Markarian~3 suggests that the Compton-reflection and the Fe K$_{\alpha}$
            emitting regions are extended to the North up to $\simeq$300~pc. The
            combination of these findings suggests that
            the optically-thick reprocessor in Markarian~3 is clumpy.}
	    {There is mounting experimental evidence for the structure of the
            optically-thick gas and dust in the nuclear environment of nearby heavily
            obscured AGN to be extended and complex.
            We discuss possible modifications to the standard unification scenarios
            encompassing this complexity. Markarian~3,
            exhibiting X-ray absorption and reprocessing on widely
            different spatial scales, is an ideal laboratory to test these models.}


\authorrunning{Guainazzi et al.}

\titlerunning{The X-ray variability history of Markarian~3}

\maketitle
%

\section{Optically thick reprocessing in AGN}

Unified scenarios
for radio-quiet Active Galactic Nuclei (AGN) (\cite{antonucci85,antonucci93}) postulate the existence of
an azimuthally-symmetric distribution of optically-thick gas and dust surrounding the
nucleus, as well as the gas clouds responsible for optical broad ($\approxgt$1000~km/s) lines.
This structure is commonly referred as the ``torus''. We will adhere to this convention in this
paper, as customary in the recent AGN literature, although
constraining the geometry of this absorbing system is one of the main astrophysical motivation
of the study presented in this paper.
Warm dust dominates the mid-IR emission in nearby AGN
(see, {\it e.g.}, \cite{alonsoherrero11})
The innermost regions of the dusty phase has to be therefore located beyond the sublimation radius
($\sim$0.4$L^{1/2}_{45}$~pc, where $L_{45}$ is the ionizing luminosity in units of 10$^{45}$~erg~s$^{-1}$;
\cite{barvainis87}).
This evidence suggests that one is dealing with a structure
at a parsec-scale distance from the active nucleus.

The unified scenarios have been shown to hold for as many as 97\% of AGN in X-ray selected samples
(\cite{mateos10}).
Violations to its basic prediction
({\it i.e.}: a different classification from optical and X-ray spectroscopy)
can be explained by
the poor quality of the spectrum on which the classification was based, variability of
the AGN emission, complexity in the absorber structure (\cite{bianchi12}),
or threshold effects on the
AGN luminosity. For instance, a minimum Eddington ratio could be necessary
for the Broad Line Regions (BLR) to form (\cite{nicastro00,marinucci12a}).

It is crucial to validate the above structure model by constraining the geometry of
the gas and dust in the AGN environment. 
Reverberation mapping of optical-UV lines against their ionizing continuum
(\cite{peterson04}) and monitoring
of changes in the column density covering highly-obscured AGN (\cite{maiolino10} and
references therein)
constrain the location of the BLR clouds within a few light-days in typical Seyfert-like
objects ({\it i.e.}, with
total bolometric luminosity $\le 10^{44}$~erg~s$^{-1}$). However, it is currently
arguable whether X-ray obscuration can be indeed ascribed to BLR clouds in most obscured
AGN, in particular in Compton-thick ones ($N_H > \sigma_t^{-1}$=$1.6 \times 10^{24}$~cm$^{-1}$).

Recently, mid-IR interferometry has allowed for the first time to effectively ``image'' the
dusty component of the parsec-scale torus. In the three cases for which such measurements
have been possible so far (NGC~1068, the Circinus Galaxy, and NGC~4151), the data are
consistent with a two-temperature distribution of dust, with most the of the obscuration
occurring on scales smaller than 1~pc (\cite{jaffe04,tristram07,burtscher09}).

Constraining the geometry of the X-ray reprocessing gas is currently
possible only by studying the response of spectral components dominated by optically-thick
reprocessing to variability of the primary continuum. These measurements are difficult, because
they require simultaneous measurements of fluxes in at least
two energy bands dominated by the AGN primary emission and by its reprocessing, respectively,
on times scales long enough to be able to detect the response of the latter to changes of the
former. These timescales are likely to be of the order of months, at least. Dedicated
monitoring campaigns of this kind are expensive, although the launch of the Swift satellite in 2004
(\cite{gehrels04}),
with its BAT X-ray transient monitoring instrument scanning continuously the high-energy
sky in the 15--150~keV band, has opened a new window in this field.

In this paper we discuss flux variability in a Compton-reflection
dominated energy band on timescales of months-years in the nearby ($z$=0.014)
prototypical Seyfert~2 galaxy Markarian~3.
Known since the dawn of X-ray astronomy, Markarian~3 hosts a
heavily obscured AGN. The most prominent feature in its X-ray spectrum is a strong iron
K$_{\alpha}$ fluorescent line (\cite{awaki91}). The continuum in the 3--7~keV range 
is dominated by unabsorbed Compton-reflection (\cite{georgantopoulos99,cappi99}).
Iwasawa et al. (1994) discovered a variability of the
hard ($E >$4~keV) luminosity by a factor of 3 in 3.6~years
by comparing their ASCA observation in April 1993 with previous
{\it Ginga} and BBXRT observations. The soft X-ray
emission had remained constant over the previous 13 years within statistics.

The paper is organized as follows: in Sect.~2 we present the observations discussed in this
paper, as well as their data reduction. The spectral analysis is described in Sect.~3, where
we also show light curves over
the first decade of this century
in energy bands dominated by different spectral components.
We discuss the implication of our results in Sect.~4, focusing on
three main aspects: the geometry of optically-thick reprocessing matter;
the location of the iron K$_{\alpha}$ line emitting gas; and the variability of the
soft X-ray emission. We summarise our
main results in Sect.~5.

In this paper: energies are shown in the source's rest reference frame; statistical errors are
calculated at the 1$\sigma$ confidence level for one interesting parameter; and the following cosmological
parameters were used to calculate luminosities:
$H_0$=70~km~s$^{-1}$~Mpc$^{-1}$, $\Lambda_0$=0.73, and $\Omega_M$=0.27 (\cite{bennett03}).
At the distance of Markarian~3, 1$\arcsec$ corresponds to 270~pc.

\section{Data reduction}

In this paper we analyse Markarian~3 spectra obtained with
imaging CCD cameras on-board Swift and XMM-Newton. The longest XMM-Newton observation (October
2000) is discussed in \cite{bianchi05}, and \cite{pounds05}.
The log of the observations discussed
in this paper is shown in Tab.~\ref{tab1}.
\begin{table*}
\caption{The log of the observations analyzed in this paper. The nomenclature ``[nm]'' indicates
the {\it Swift}/XRT observations which have been merged together. For instance: ``0003546000[23]''
means that a single spectrum was extracted from Obs.\#00035460002 and Obs.\#00035460003.}
\label{tab1}
\begin{center}
\begin{tabular}{lcccc} \hline \hline
Obs.\# & Start Time & $T_{exp}$ & Instrument & Mode \\ 
& & (ks) & & \\ \\ \hline
0111220201 & 2000-10-19T15:39:46 & 37.0 & EPN & PrimeFullWindow \\
0009220601 & 2001-03-20T20:49:05 & 4.9 & EPN & PrimeFullWindow \\
0009220701 & 2001-03-28T21:33:43 & 3.0 & EPN & PrimeFullWindow \\
0009220901 & 2001-09-12T01:12:44 & 1.7 & EPN & PrimeFullWindow \\
0009221601 & 2002-09-16T05:32:03 & 1.6 & EPN & PrimeFullWindow \\
0009220401 & 2002-03-10T13:57:37 & 2.6 & EPN & PrimeFullWindow \\
0009220501 & 2002-03-25T17:45:25 & 3.4 & EPN & PrimeFullWindow \\
00035460001 & 2006-03-21T07:16:01 & 5.1 & XRT & PHOTON \\
0003546000[23] & 2006-04-02T12:00:00 & 18.1 & XRT & PHOTON  \\
00037226001 & 2008-01-22T00:40:08 & 9.4 & XRT & PHOTON \\
0003546000[45] & 2012-01-11T12:00:00 & 14.7 & XRT & PHOTON \\
0656580301 & 2012-03-15T12:32:16 & 30.9 & EPN & FastTiming \\
00035460008 & 2012-05-06T03:18:12 & 9.2 & XRT & PHOTON \\
\hline \hline
\end{tabular}
\end{center}
\end{table*}

\subsection{XMM-Newton/EPIC-pn}

In this paper we discuss only spectra extracted with the EPIC-pn camera (\cite{struder01}).
Markarian~3 was
observed seven times in the early phase of the XMM-Newton mission using the Full Frame Mode.
The March
2012 observation was instead performed in Timing Mode. Raw scientific telemetry (Observation
Data Files)
was reduced using SAS Version 12 (\cite{gabriel03}), and the calibration files available on
June 1 2012.
Spectra were extracted from calibrated event lists generated with the task {\tt epchain}. The
Timing
Mode event list energy scale was corrected for any rate-dependent effect, which is, however,
likely to
be negligible due to the source flux level.

For imaging modes,
source spectra
were extracted using circular regions of 40$\arcsec$ radius
centered around Markarian~3 optical coordinates
($\alpha_{2000}$=93.9015194$^{\circ}$; $\delta_{2000}$=71.0375254$^{\circ}$); background spectra were
extracted from circular regions free of serendipitous contaminating sources on the same CCD quadrant as
the boresight position, and corresponding to {\tt RAWY} positions close to where Markarian~3 is located
on the EPIC-pn boresight CCD, to ensure that
a similar Charge Transfer Efficiency correction
applies.

For the Timing Mode observation, source spectra were extracted from a box in
detector coordinates
centered on {\tt RAWX}=36 and with a radius of 5 pixels (each pixel in the EPIC-pn camera
corresponds to
4.4$\arcsec$). No constraints were applied to the pseudo-spatial coordinate {\tt RAWY},
which
represents the elapsed time during the observation in this mode. Background spectra were extracted
from a nearby
box centered on {\tt RAWX}=50 to ensure that the background spectrum represent as faithfully
as possible
the true background underneath the source position. The background represents about 75\%
of the source
level in the 4--10~keV energy band. The emission profile along the {\tt RAWX} coordinate
shows a gradual
decrease (once the signal corresponding to Markarian~3 is removed)
by about 5\%
between {\tt RAWX}=27 and {\tt RAWX}=55. Through a linear interpolation to {\tt RAWX}=36,
we estimate that our background spectrum could underestimate the true background level
underneath the
source by about 3\%. We have added this factor in the analysis of the Timing Mode spectrum.

Response files appropriate for each observation were generated with the SAS tasks {\tt rmfgen}
and
{\tt arfgen}. The EPIC-pn response is known to be stable within 3\% (\cite{sartore12}).

\subsection{Swift/XRT}

Swift/XRT (\cite{burrows05}) observed Markarian~3 on 9 occasions. However, as 2 observations have
relatively short exposure time (2~ks and 55~s respectively), we consider
here 7 Swift XRT observations with exposure times longer than 5~ks.
We extracted Swift XRT products from the cleaned photon
counting event files following the procedures described in {\it The
SWIFT XRT Data reduction Guide
V.1.2}\footnote{http://swift.gsfc.nasa.gov/docs/swift/analysis/}. Spectra
of Markarian~3 were extracted from circular
regions of radius 60$\arcsec$, while background spectra were obtained from
larger (100$\arcsec$ in radius) source--free regions close to the source. We
used the standard response files for the photon counting mode with
grades 0 to 12 which are part of the current Swift XRT calibration
database, taking into account the observation epoch. Ancillary files
have instead be created for each observation using the {\tt xrtmkarf} task
for the specific source position. 
After checking that the individual spectra were
consistent, we have merged the spectra from ObsID 00035460002 and
00035460003 (3 days apart) and from ObsID 00035460004 and 00035460005
(1 day apart). Hence 5 X--ray XRT spectra of of Markarian~3 are considered
here.

\subsection{Swift/BAT}

The Swift-BAT (\cite{barthelmy05})
data were processed with the {\tt Bat\_Imager} code (\cite{segreto10}).
The 15-150 keV light curve, as well as those in three energy bands: 15--30, 30--70, and 70--150~keV,
were obtained producing a set of
all-sky maps each covering 15 days, starting from December 2005 up to
March 2012, and extracting the count rate and its error from the pixel
corresponding to the source position in each of the maps.

\section{Spectral analysis}
\label{sect_analysis}

All the EPIC-pn and XRT spectra were independently analysed in the nominal 4--10~keV band,
where Compton-reflection dominates (\cite{iwasawa94}). However, the upper bound is
observation-dependent (see Fig.~\ref{fig7}) as XRT dost not confidently detect the source
up to 10~keV in most observations.
Background-subtracted
spectra were fitted using the
Cash statistics ($C$; \cite{cash76}) for us to be able to use the natural instrumental spectral binning.

Our first goal was to measure the light curves
of the fluxes of spectral components associated with reflection of
the
primary continuum by optically thick matter. We have therefore fit
the spectra with the following phenomenological model, $M_{pr}$ (``pure reflection''):
$$
M_{pr} = C_R(\Gamma,E) + \Sigma_{i=1}^3 G_i(E) 
$$
where $C_R$ is a Compton-reflection continuum component (\cite{magdziarz95}) and the $G_i$ are unresolved
({\it i.e.}, intrinsic width assumed to be equal to 0)
Gaussian emission lines. In the Compton-reflection component we have assumed: no high-energy
cut-off of the primary continuum (as this will most likely fall outside the EPIC-pn and XRT
sensitive bandpass: \cite{dadina07}); solar abundances (according to Anders \& Grevesse 1983);
and a 45$^{\circ}$ inclination angle between
the normal to the plane of the Compton-reflecting slab and the line-of-sight.
The only free parameter in this component, apart from its normalization, is therefore
the photon index of the power-law primary continuum, $\Gamma$.
The centroid energies of the Gaussian profiles were fixed to 6.4, 6.7, and 6.96~keV,
respectively,
corresponding to
fluorescent emission from neutral or mildly ionized iron, the resonant component of
the He-like triplet and the H-like iron line, respectively.
Sako et al. (2000) report for the He$\alpha$ an energy of 6.685~keV (without error)
from their analysis of the {\it Chandra}/HETG spectrum of Markarian~3.
This energy is intermediate between that of the resonant, the intercombination (6.668--6.682~keV)
and the forbidden (6.637~keV) component. Neither the EPIC-pn nor the Swift/XRT spectra can
constrain this parameter. We have therefore opted for fixing the
centroid energy to the laboratory value of the resonant component following the results
after Bianchi et al. (2005), which rule out an important contribution
of the forbidden component. The results presented in this paper are
insensitive to this choice.
The intensity of the He-like and H-like lines can be constrained only in the longest
XMM-Newton observation:
$I_{6.7}$=(7.0$\pm$1.4)$\times$10$^{-6}$ and
$I_{6.96}$=(5.5$\pm$1.4)$\times$10$^{-6}$~photons~cm$^{-2}$~s$^{-1}$,
respectively.
In the other observations they have been fixed to the above best-fit values. This assumption
is astrophysically
justified, because a substantial fraction of them
is likely to produced in the X-ray Narrow Line Regions, extended
on scales of the order of a few hundred parsecs (see Bianchi et al. 2006 for a quantitative
discussion of this point).
$M_{pr}$ does not leave unaccounted for residuals in any of the spectrum.
The best-fit parameter values are reported in. Tab.~\ref{tab2}.
\begin{table*}
\caption{Best-fit values of the free parameters in $M_{pr}$ ($\Gamma$, $F_{4-5}$, $I_{6.4}$, $C/\nu$)
and of the phenomenological model used to fit the 0.5--2~keV energy band ($\Gamma_s$, $F_{0.5-1.5}$).
$F_{4-5}$ ($F_{0.5-1.5}$) is the flux of the Compton-reflection component (total flux)
in the 4--5
(0.5--1.5)~keV energy range; 
$I_{6.4}$ is the normalization of the iron K$_{\alpha}$ line; $C$ is the best-value Cash
statistic value.}
\label{tab2}
\begin{center}
\begin{tabular}{lcccccc} \hline \hline
& $\Gamma$ & $\Gamma_s$ & $F_{0.5-1.5}$$^a$ & $F_{4-5}$$^a$ & $I_{6.4}$$^b$ & $C/\nu$ \\ 
& & & & & \\ \\ \hline
0111220201 & $1.00 \pm 0.04$ & $2.79 \pm 0.03$ & $-12.112 \pm 0.004$ & $-12.38 \pm 0.01$ & $4.07 \pm 0.17$ &  1338.6/1289 \\ 
0009220601 & $1.19 \pm 0.014$ & $2.85 \pm 0.08$ & $-12.117 \pm 0.011$ & $-12.48 \pm 0.04$ & $4.0 \pm 0.5$ &  575.9/769  \\ 
0009220701 & $1.2 \pm 0.2$ & $2.78 \pm 0.11$ & $-12.109 \pm 0.014$ & $-12.55 \pm 0.06$ & $4.2 \pm 0.6$ &  367.0/533  \\ 
0009220901 & $1.1 \pm 0.3$ & $3.14 \pm 0.15$ & $-12.110 \pm 0.019$ & $-12.53 \pm 0.08$ & $3.2 \pm 0.8$ &  248.3/368  \\ 
0009221601 & $1.4 \pm 0.3$ & $3.17 \pm 0.15$ & $-12.090 \pm 0.019$ & $-12.58 \pm 0.08$ & $3.9 \pm 1.0$ &  239.7/326  \\ 
0009220401 & $1.3 \pm 0.2$ & $2.89 \pm 0.12$ & $-12.121 \pm 0.015$ & $-12.51 \pm 0.06$ & $3.4 \pm 0.7$ &  404.0/520  \\ 
0009220501 & $1.5 \pm 0.2$ & $2.99 \pm 0.10$ & $-12.100 \pm 0.013$ & $-12.52 \pm 0.05$ & $3.3 \pm 0.5$ &  409.2/580  \\ 
00035460001 & $1.1 \pm 0.8$ & $2.8 \pm 0.4$ & $-12.18 \pm 0.05$ & $-12.42 \pm 0.11 $ & $3.4 \pm 1.7$ & 70.5/86  \\ 
0003546000[23]& $1.7 \pm 0.3$ & $2.60 \pm 0.16$ & $-12.14 \pm 0.02$ & $-12.37 \pm 0.05 $ & $2.5 \pm 0.8$ & 213.6/259  \\ 
00037226001 & $0.8 \pm 0.8$ & $2.8 \pm 0.3$ & $-12.25 \pm 0.04$ & $-12.66 \pm 0.11$ & $3.1 \pm 1.2$ &  78.4/102  \\ 
0003546000[45] & $1.4 \pm 0.3$ & $2.76 \pm 0.14$ & $-12.16 \pm 0.03$ & $-12.37 \pm 0.06$ & $5.1 \pm 1.7$ &  202.2/242  \\ 
0656580301 & $0.8 \pm 0.3$ & $3.02 \pm 0.10$ & $-12.163 \pm 0.010$ & $-12.54 \pm  0.08$ & $4.0 \pm 0.5$ &  1314.8/1296  \\
00035460008 & $1.4 \pm 0.5$ & $2.4 \pm 0.2$ & $-12.25 \pm 0.03$ & $-12.53 \pm 0.09$ & $3.3 \pm 1.3$ &  95.2/127  \\  
\hline \hline
\end{tabular}

\noindent
$^a$base-10 logarithm of the flux in units of erg~cm$^{-2}$~s$^{-1}$

\noindent
$^b$in units of 10$^5$~photons~cm$^{-2}$~s$^{-1}$

\end{center}
\end{table*}

\begin{figure*}[ht]
\begin{center}
\includegraphics[height=135mm,angle=90]{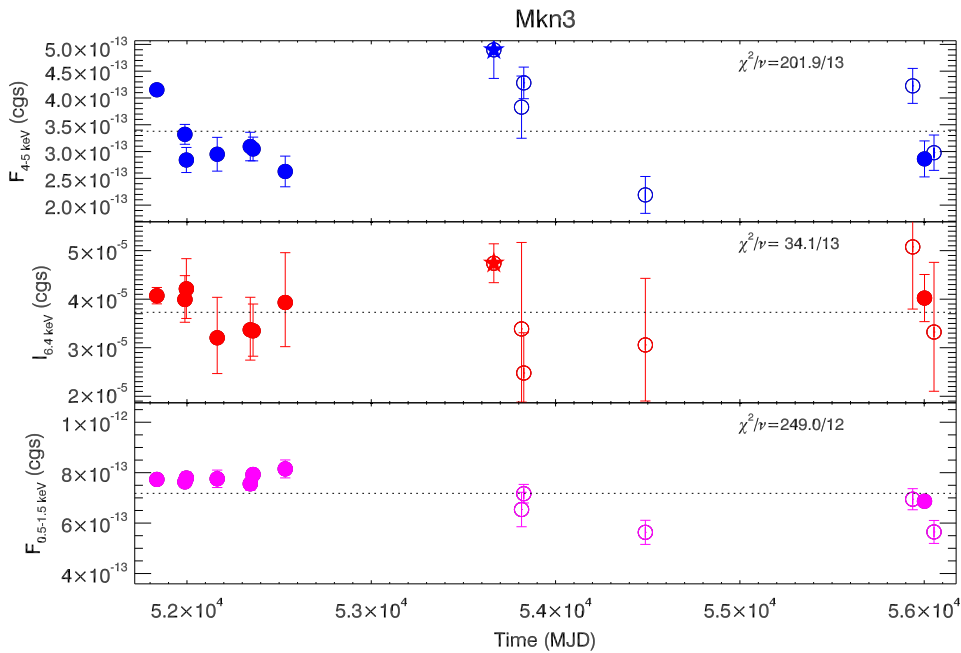}
\end{center}
\caption{
Light curves of the: observed flux in the 4--5~keV energy band ({\it upper panel});
intensity of the Fe K$_{\alpha}$ line ({\it medium panel}); observed flux
in the 0.5--1.5 keV energy band ({\it lower panel}). The {\it numbers in each
panel} indicate the chi-squared value for a ft against a constant versus the
number of degrees of freedom. {\it Filled circles}: XMM-Newton/EPIC; {\it empty circles}:
Swift/XRT; {\it filled star}: Suzaku. The scale on the y-axis in all panels corresponds
to $\pm$50\% of the mean level ({\it dotted lines}).
}
\label{fig1}
\end{figure*}

\begin{figure}
\includegraphics[height=90mm,angle=90]{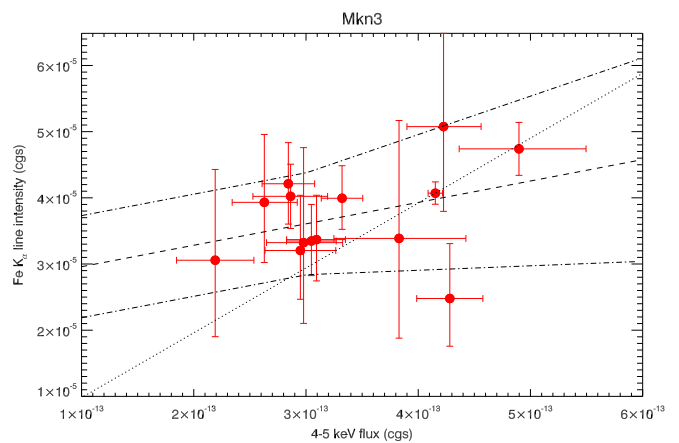}
\caption{
Intensity of the Fe K$_{\alpha}$ against the observed flux in the 4--5~keV energy band.
The {\it dotted line} indicates the locus of constant line EW, normalized to the
value measured during the deepest XMM-Newton observation (October 2000). The
{\it dashed line} indicates the linear best-fit, whereas the {\it dot-dashed lines}
indicate the envelope corresponding to the 1$\sigma$ uncertainties on the best-fit
parameters.
}
\label{fig2}
\end{figure}

\begin{figure}
\includegraphics[height=90mm,angle=90]{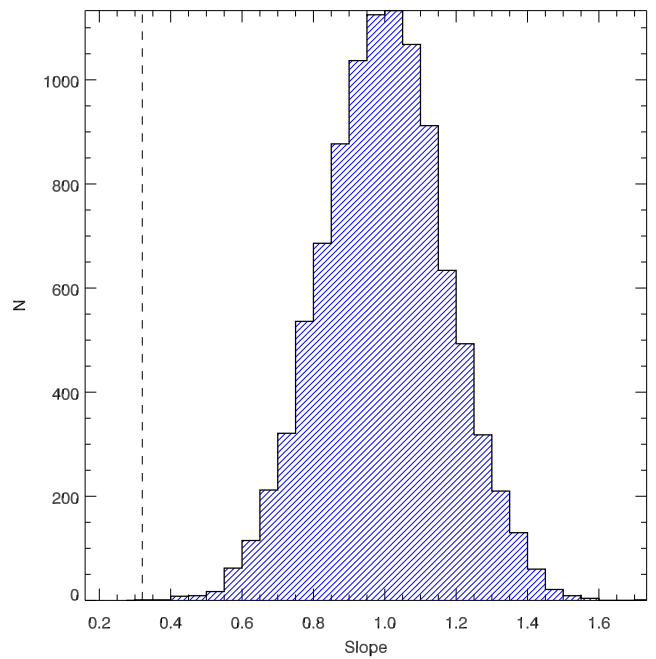}
\caption{
Distribution of slopes of the best liner fit between the 4--5~keV Compton fluxes and
Fe line intensities, when the latter are simulated assumption that they responded
simultaneously to the same primary emission. The {\it vertical line}
indicates the measured slope $\hat b$.
}
\label{fig11}
\end{figure}

Fig.~\ref{fig1} ({\it upper} and {\it medium} panels) shows the light curves of the
observed flux in the
4--5~keV band, and of the intensity of the K$_{\alpha}$ fluorescent iron line.
We have also added the results of the {\it Suzaku} pointed observation of Markarian~3
discussed by Awaki et al. (2008; Obs.\#1000040010)\footnote{Markarian~3 was observed also
by the {\it Chandra}/HETG in combination with ACIS-S
exactly seven months before the first XMM-Newton observation. We
downloaded the spectrum from the {\tt TGCat} archive, and analyzed it using the same
procedure and model as the EPIC-pn and XRT spectra. While the intensity of the iron
K$_{\alpha}$ line is comparatively well constrained
[$I_{6.4}$=$(5.0 \pm 0.5) \times 10^{-5}$~erg~cm$^2$~s$^{-1}$; \cite{sako00}],
the 4--5~keV Compton flux is
not ($\log_{10}(F_{4-5})$=$-12.8 \pm 0.7$; cf. Tab.~\ref{tab2}). We will therefore not discuss
these data in our paper.}
We decided not to analyze again the XIS spectra, because they
were reduced and published
by a highly experienced team, and no significant change in the calibration
occurred above 3~keV since the time of the original data reduction in Awaki et al. (2008).
The 4--10~keV energy band, chosen because it is dominated by emission-line free Compton-reflection,
shows a variability dynamical range larger then 70\%. Similar dynamical
ranges are seen when comparing the XMM-Newton or Swift observations among each other,
ruling out cross-calibration effects (which should be anyhow of the order of at most
10\%; \cite{tsujimoto11}). The chi-squared for a fit against a constant, $\chi^2_c/\nu$,
is 199.9/13. On the other hand, the iron line light curve exhibits
a lower variability
dynamical range. The $\chi^2_c/\nu$ value is 34.1/13.
The two quantities are very weakly correlated (Fig.~\ref{fig2}). 
A linear fit $I/\langle I \rangle = \hat a + \hat b \times F/\langle F \rangle$
yields\footnote{We normalize here $I$ and $F$ to their mean
in order to reduce the
errors on the best-fit parameters, which
otherwise could be artificially increased
by the leverage induced by the large difference in absolute value
between the fitted quantities}: $\hat a = 0.7 \pm 0.2$
and $\hat b = 0.32 \pm 0.17$. The two-side
significance of the Spearman's rank correlation coefficient is $\simeq$31\% only.
This indicates that
Equivalent Width (EW) is not constant. However, care must be exercised in interpreting this
result, as well as the low $\chi^2_c/\nu$ of the Fe K$_{\alpha}$ intensity light curve, because
the statistical uncertainties on its measurements are a factor of 2--3 larger
than those on the 4--5~keV Compton flux.
To make a more quantitative assessment of the correlation likelihood 
(or lack thereof) between the observables in Fig.~\ref{fig2}, we
performed Monte-Carlo simulations of 
the Fe K$_{\alpha}$ intensity light curve, assuming that it follows the same relative trend
(in units of ratio against the mean) as the measurements of the 4--5~keV flux.
In astrophysical terms,
the simulated scenario assumes that
the continuum Compton-reflection and the Fe line intensity
simultaneously respond to changes of the same primary continuum.
For each observation, 10$^4$ realizations of a Gaussian distribution
where drawn from a random sample, whose mean is
the Fe line intensity aligned to the Compton-flux,
and whose standard deviation is the measured statistical uncertainty on the Fe line intensity.
For each simulated light curve, we calculated the slope of the best linear fit between
the Compton flux and the simulated Fe line intensity. In Fig.~\ref{fig11} we compare the
distribution of the simulated slopes
with the measured slope. The probability of getting a value of $b \le \hat b$ is
$\le$10$^{-4}$.

Variation of the flux in the 4--5~keV energy band could be due to a change of the
Compton-reflected flux in this band, as well as to the emerging of different spectral
components. It is in principle possible, for instance, that a reduction
of the photoelectric column density
covering the primary AGN emission brings a fraction of the primary emission into this
energy band during some of the observations presented in Fig.~\ref{fig1}.
The intrinsic spectral indices yielded by $M_{pr}$ are significantly flatter than
typically observed in Seyfert galaxies (\cite{bianchi09a}) and measured in X-ray
broad-band observations of Markarian~3 (\cite{cappi99,awaki08}). While we warn against
interpreting $\Gamma$ in $M_{pr}$ as a reliable measurement of the photon index of the
primary continuum on spectra with a small sensitive bandpass, the systematic flat
spectral indices
may indicate that a different spectral model is required to properly fit the data.

In order to test the impact that different origins
for the spectral variability may have on our
measurements of the Compton-reflection
variability, we fit the EPIC-pn and XRT spectra with two alternative models.
These models have been applied to the EPIC-pn and XRT spectra discussed in this
paper under the same conditions as $M_{pr}$. In particular, we fit the data in the same
energy band: 4--10~keV. Because we want to address the
flat spectral indices yielded by $M_{pr}$, in the reminder of this Section we will fix 
the value of this parameter to that determined by Awaki et al.
(2008; $\Gamma_{int} \equiv 1.8$) from their
analysis of the {\it Suzaku} observation of Markarian~3, unless otherwise
specified. This choice is justified by the
fact that the spectral coverage of the scientific payload on board {\it Suzaku} up to
several tens of keV allows one to better constrain the intrinsic shape of the
primary continuum in such an obscured source.

\noindent
- \underline{unobscured reflection plus obscured transmission}. We assumed the full continuum
model as discussed in \cite{bianchi05}:
$$
M_{urot} = C_R(\Gamma_{int},E) + \Sigma_{i=1}^3 G_i(E) + e^{-\sigma_{ph}(E) N_{H,p}} \times AE^{\Gamma_{int}}
$$
There is a strong correlation between the primary
spectral index and the iron abundance of the Compton-reflection component
(see, {\it e.g.}, the discussion in Bianchi et al. 2005), primarily
driven by the depth of the neutral iron photo-absorption edge. We therefore imposed
$Z_{Fe}$ to vary within the confidence interval as determined from the deepest XMM-Newton
observation: $Z_{Fe}$=0.43$\pm$0.09$Z_{\odot}$.
The photoelectric
cross-section $\sigma_{ph}(E)$ is according to the T\"ubingen-Boulder model with the
associated abundances (\cite{wilms00}), and includes also
the contribution from optically-thin Compton-scattering\footnote{We employed in this context
the implementation {\tt cabs} in {\sc Xspec} for comparison with previous works,
despite the issues described in Murphy \& Yaqoob (2009)
which should discourage its use for Compton-thick absorbers. They
primarily affect the determination of the absorption-corrected luminosity
of the obscured primary, a measurements in which we are not interested for the scope
of this paper.}
The typical column densities measured in the EPIC-pn spectra are
$N_{H,p}$$\simeq$1.2--1.6$\times$10$^{24}$~cm$^{-2}$. In the deepest EPIC-pn
observations:$N_{H,p}$=$(1.19 \pm 0.06)$, and $(0.49 \pm^{0.10}_{0.05}) \times 10^{24}$~cm$^{-2}$
in October 2000 and March 2012, respectively. Lower limits from the XRT spectra
are in the range $N_{H,p}$$\simeq$0.6--1.4$\times 10^{24}$~cm$^{-2}$.
We urge readers not to over-interpret the measurements of the absorber
column density in this scenario.
The column density of the absorber cannot be constrained by our data if the value of
the photon index is left free to vary.
Fixing the shape of the spectral index severely limits the size of the confidence interval
parameter space.

\noindent
- \underline{obscured reflection}. We added a photoelectric absorption component
covering the Compton-reflection emission:
$$
M_{or} = e^{-\sigma_{ph}(E) N_{H,r}} C_R(\Gamma_{int},E) + \Sigma_{i=1}^3 G_i(E)
$$
We again constrained $Z_{Fe}$=0.43$\pm$0.09$Z_{\odot}$.
In astrophysical
terms, $M_{or}$ represents a scenario whereby the line-of-sight to the inner far side
of the torus grazes the rim of its near side or crosses a
dust lane in the host galaxy (\cite{lamastra08}). Although
gas from the torus rim is likely to be evaporated by the AGN radiation pressure,
and therefore significantly ionized, we used in $M_{or}$
a photoelectric model for cold matter, as the data do not require the ionization parameter
as a further degree of freedom. We refrain from deriving any
astrophysical inference from the measured column densities,
due to the simplicity of the assumptions and to the statistical
quality of most of the spectra analyzed in this paper.
Not surprisingly, only the two deepest XMM-Newton observations
require a column density significantly different from zero: $N_{h,r}$=$4.5 \pm 0.6$ and
$(17 \pm^6_2) \times 10^{22}$~cm$^{-2}$ in October 2000, and March 2012, respectively. In the other
observations, the statistical quality of the spectra is the main driver behind
upper limits in the range 1.7--6$\times$$10^{22}$~cm$^{-2}$.

\begin{figure}
\includegraphics[height=90mm,angle=-90]{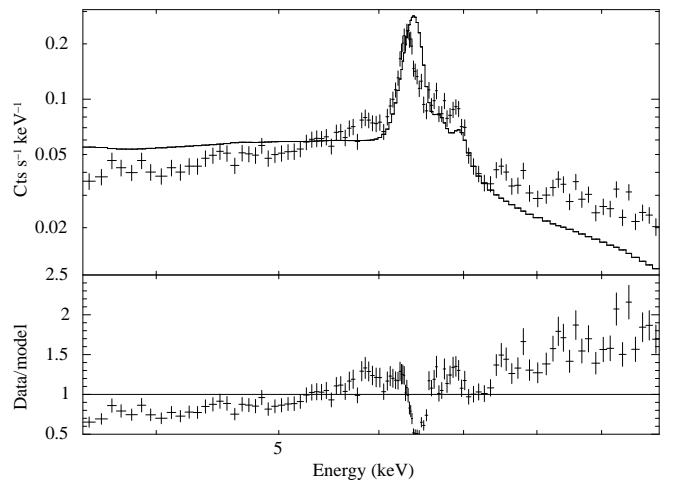}
\caption{
Spectrum of the XMM-Newton October 2000 observations ({\it upper panel}) and
residuals in units of data/model ratio ({\it upper panel}) if a model
of ionized reflection with $\Gamma_{int}$=1.8 is applied.
}
\label{fig12}
\end{figure}

\noindent
- \underline{ionized reflection}. We replaced the Compton-reflection component in $M_{pr}$ with
a model explicitly including the ionization state of the reflector\footnote{through the
ionization parameter $\xi \equiv L/(nR^2)$, where
$L$ is the ionizing luminosity, $n$ is the
electron density, and $R$ the distance between the reflector and the source of
ionizing continuum}, producing also the ionized iron lines which in other models
are fit empirically with Gaussian profiles.
This model is
based on the calculation by Ross \& Fabian (2005). If we assume $\Gamma_{int}$=1.8,
the model does not yield a good fit to the best-quality spectra (see, {\it e.g.}, the
results on the deepest XMM-Newton observation in Fig.~\ref{fig12}).
If left free to vary, the spectral
index pegs to the minimum value allowed by the model (1.6), still leaving significant residuals.
Even
when applied on spectra with a lower statistical quality,
where the formal quality of the fit is comparable to that yielded by $M_{pr}$,
the model exhibits:
a) too broad an iron line profile, and; b) too shall a photoelectric
absorption edge with respect to the data.
We will not consider therefore this model any longer in this paper.

\begin{figure}
\includegraphics[height=90mm,angle=90]{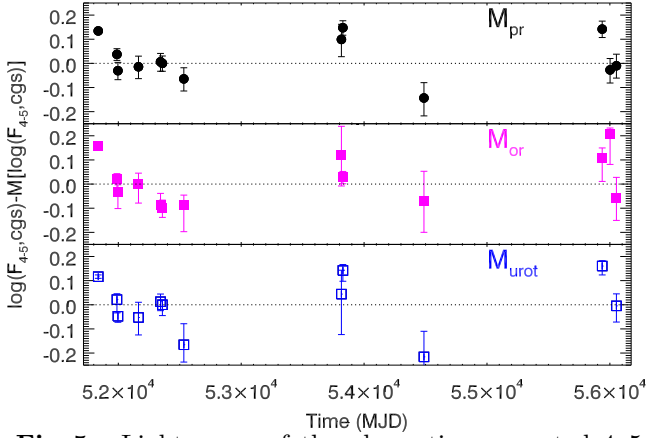}
\caption{
Light-curve of the absorption-corrected 4--5~keV flux
of the Compton-reflection component when models
$M_{pr}$ ({\it upper panel}; the same quantity as in
the {\it upper panel} of Fig.~\ref{fig1}), $M_{or}$ ({\it medium panel}),
and $M_{urot}$ ({\it lower panel}) are used. The units on the y-axis are
the difference between the logarithm of the flux and its median.
The {\it dotted line} indicates the median level.
}
\label{fig10}
\end{figure}

In Fig.~\ref{fig10} we compare the light curves of the 4--5~keV flux due solely to the
Compton-reflection component as measured in different spectral models, and corrected for
the intervening absorption in the $M_{or}$ model. Differences are apparent.
The most evident is that the error
bars in $M_{or}$ and $M_{urot}$ are larger, due to the larger number of degrees of freedom.
However, the overall Compton-reflection variability
pattern is not qualitatively modified. The main
change is the measured flux of the March 2012 EPIC-pn observation. It is larger
by about 0.1~dex with $M_{or}$ when compared to $M_{pr}$,
and is fully unconstrained with $M_{urot}$.

In summary: the fact that the
flux associated to the Compton-reflection component varies {\it does not depend} on the
spectral model used to fit the data analyzed in this paper (within the simple spectral models used).
On the other hands, quantitative estimates
of the variability pattern timescales {\it do} depend on the spectral modeling.
Reader should bear in mind this important caveat over the rest of the paper.
Long-term deeper monitoring of Markarian~3
would be required to break this degeneracy, where each individual spectrum has got a
statistical quality comparable or better than the deepest EPIC-pn observation discussed
in this paper.

Finally,
in the {\it lower panel} of Fig.~\ref{fig1} we show the light curve in the 0.5--1.5~keV
energy band.
It was measured by fitting the spectra in the 0.5--2~keV with a phenomenological model
constituted by a power-law (of photon index $\Gamma_s$)
and 7 unresolved Gaussian profiles, with energy and normalization fixed to
the best-fit values obtained from the deepest XMM-Newton observations.
We did not include the {\it Suzaku} measurement in this panel, because it is likely to
be contaminated by the ROSAT source IXO30 (\cite{bianchi05,awaki08}).
Variability with a dynamical range of about $\pm$10\% is seen also in this band
($\chi^2_c/\nu$=249.0/12). Such a dynamical range is of the same order of the systematic
uncertainties in the cross-calibration among flying X-ray detectors in this energy range
(\cite{tsujimoto11}). This casts doubts on an astrophysical origin of these differences.
We discuss this finding in more details in Sect.~\ref{sect_soft} of this paper.

\section{Variability timescales in Markarian~3}
\label{sect_times}

\begin{figure*}[ht]
\begin{center}
\includegraphics[height=135mm,angle=90]{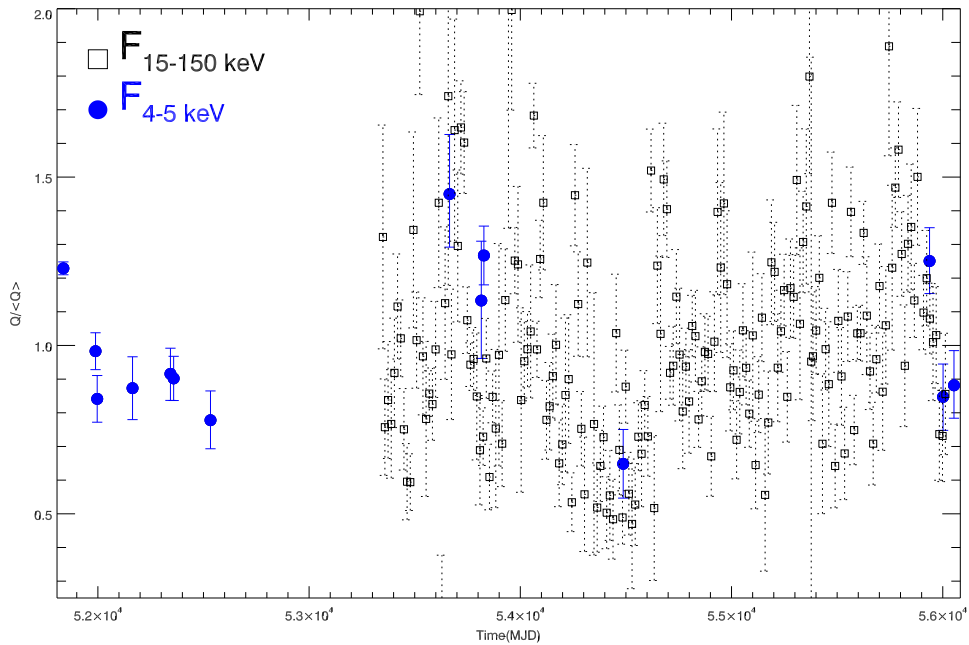}
\end{center}
\caption{
Light curve in the 4--5~keV energy band ({\it filled circles}) against the Swift/BAT
light curve in the 15--150~keV energy band ({\it empty squares}). Both curves are
normalized to their respective mean for direct comparison.
}
\label{fig4}
\end{figure*}

In Fig.~\ref{fig4} we compare the 4--5~keV observed flux against the flux measured by
the Swift/BAT in the 14-195~keV energy band. Both curves are normalized to their respective
mean. The 4--5~keV points are calculated with the $M_{pr}$ model.

\begin{figure}
\includegraphics[height=90mm,angle=90]{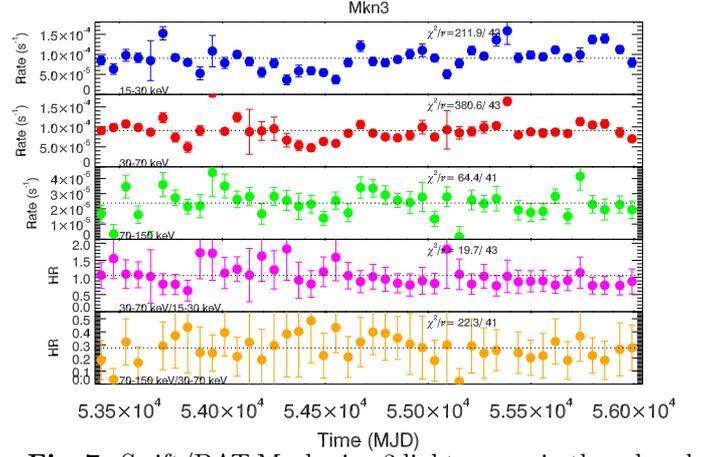}
\caption{
Swift/BAT Markarian~3 light curves in three bands, and their hardness ratio.
}
\label{fig6}
\end{figure}

The variability observed in the BAT light curve could be due either to true changes
of the intrinsic nuclear flux
or to changes of the column densities covering the source of the primary emission.
In the latter case we should see different variability patterns in light
curves
extracted in different BAT energy ranges, with the light curve extracted from the hardest band
tracing most closely the primary continuum. In Fig.~\ref{fig6} we show the Markarian~3 BAT
light curves
in three bands: 15--30~keV; 30--70~keV, and 70--150~keV, together with their hardness ratios.
They exhibit a qualitatively similar pattern. The hardness
ratios are consistent with being constant, with $\chi^2_c/\nu$ values of 19.7/43
and 22.3/41 for the 30--70/15--30 and the 70--150/30--70~keV hardness ratio, respectively.

With the light curves shown in Fig.~\ref{fig4} and \ref{fig6} we can estimate the following
timescales:

\noindent
- {\it The minimum variability timescale of the primary emission, $\tau_p$}. We define
it from the shortest interval between two bins of the BAT light curve which are inconsistent
at the 95\% confidence level. This quantity is basically constrained by the
statistics of the
BAT light curve.
There are several consecutive
bins, both in the full as well as in the energy-resolved BAT light curves,
which are inconsistent at this statistical level. We conclude that
$\tau_p \approxlt$15~days. Significantly shorter (hours to days) timescales
are common in Seyfert galaxy (\cite{barr86,mchardy06}), but they are basically inaccessible to us
in such heavily obscured source as Markarian~3

\noindent
- {\it The maximum variability timescale of the Compton flux, $\tau$}. We define it as the
minimum time interval between two bins of the Compton flux light curve differing at more than the
95\% confidence level. It depends on the spectral model, due to the larger statistical
errors affecting the measurement of the Compton flux where more complex spectral models are used
(Tab.~\ref{tab3})

\noindent
- {\it The delay in the response of the Compton flux to variation of the primary
continuum $\Delta t_s$}
We calculate the quantity:

$$
\chi^2 (\Delta t_s) = \Sigma [ (F^n_{4-5}(\Delta t_s)-F^n_{14-195})/(\sigma^2_{4-5}+\sigma^2_{14-195}) ]
$$
where: $F^n_{\Delta E}$ are the normalized fluxes in the $\Delta E$ energy range; $\sigma_{\Delta E}$
are the corresponding statistical uncertainties, augmented by
5\% to take into account possible systematic cross-calibration
uncertainties; and $\Delta t_s$ is a temporal shift applied
to the 4--5~keV flux data points only.
For this cross-correlation to be astrophysically
meaningful (cf. Sect.~\ref{sect_scales}),
the BAT light curve must be binned with $\Delta t_{BAT} > \tau$.
We used the smaller integer multiple of 15~days fulfilling this relation.

\begin{figure}
\includegraphics[height=90mm,angle=90]{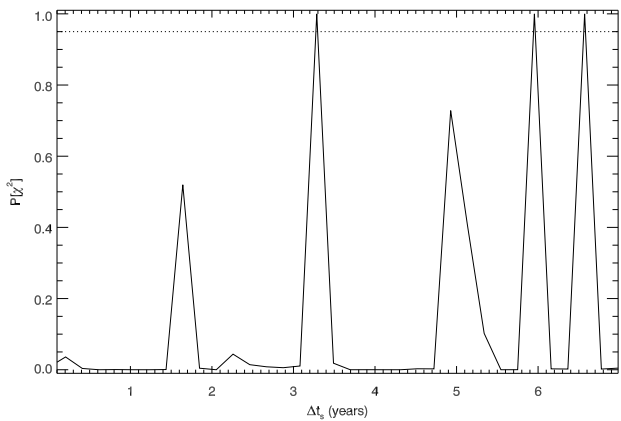}
\caption{Probability of $\chi^2 (\Delta t_s)$ as a function of $\Delta t_s$ for $M_{pr}$ when
$\Delta t_{BAT}$~=~75~days. The {\it dotted line} indicates the 95\% confidence level.
}
\label{fig14}
\end{figure}

We consider this
cross-correlation statistically significant
when the one-sided probability of $\chi^2 (\Delta t_s)$ is larger then 95\%. For $M_{pr}$,
a family of solutions exists, corresponding to $\Delta t_s \approxgt 1200$~days (see
Fig.~\ref{fig14}). If
the 4--5~keV Compton flux is calculated using the other spectral models,
no $\Delta t_s$ satisfies our statistical criterion.
\begin{table}
\caption{Minimum variability timescale of the Compton-reflection component, $\tau$, and
delay in the response of the Compton-reflection to changes of the primary continuum, $\Delta
t_s$. All quantities
are in days. Empty cells indicate unconstrained measurements.}
\label{tab3}
\begin{tabular}{lcc} \hline \hline
Model & $\tau$ & $\Delta t_s$  \\ \hline
$M_{pr}$ & $64.0 \pm 0.3$$^a$ & $\ge 1162.5$ \\
$M_{or}$ & $152.2 \pm 0.2$$^b$ & ... \\
$M_{urot}$ & $115.64 \pm 0.13$$^c$ & ... \\ \hline \hline
\end{tabular}

\noindent
$^a$time difference between the 2012-01-11 and the 2012-03-15 observations.

\noindent
$^b$time difference between the 2000-10-19 and the 2001-03-20 observations.

\noindent
$^c$time difference between the 2012-01-11 and the 2012-06-05 observations.

\end{table}

In principle, the variable Compton-reflection component could contribute to the
variability of the BAT light curves as well. While spectral decompositions in continuum
components
must been taken with a grain of salt when performed in spectral regions where the instrumental
resolution is poor, the results of the {\it Suzaku} observation
formally indicate that the Compton-reflection
could contribute about one third of the total flux at 15~keV, and about 50\%
thereof above 30~keV (cf.
Fig.2 in \cite{awaki08}). However, it is unlikely that the Compton-reflection
contribute significantly to
the variability above 15~keV on the following grounds: a) the short variability
timescale of the BAT light curve; b) its energy invariance: Compton-reflection and primary
continuum exhibit very different spectral shape around the Compton peak at 30~keV; c) the
fact that there is very little correlation between the 4--5~keV and the 15--150~keV light
curve at zero time shift.
The probability of $\chi^2(\Delta t_s=0)$ is $\sim$6$\times 10^{-4}$ for $M_{pr}$.

\section{Discussion}

\subsection{Constraining the torus geometry through variability}
\label{sect_scales}

In principle the variability timescales determined in Sect.~\ref{sect_times} can be used
to constrain the location and geometry of the optically-thick reprocessor. We discuss in
this Section the method, its assumptions, and its limitations, as well as the results of
its application to the monitoring campaign of Markarian~3 discussed in this paper.

If the primary continuum illuminating a slab of optically thick gas
varies on a timescale $\tau_p$, the variability of the primary emission can be fully
diluted if the light crossing time across the reflecting part of the slab is larger
than $\tau_p$. For all the spectral model discussed in this paper $\tau \gg \tau_p$ (Tab.~\ref{tab3}).
Due to its finite size, the reprocessor will not respond to the instantaneous primary flux,
rather on
an average over its own light-crossing time. This is the reason why we impose
in our cross-correlation analysis (Sect.~\ref{sect_times}) the condition $\Delta t_{BAT} > \tau$.

Under this condition, the delayed response of the reflected component to changes of the primary continuum
can be used to constrain the distance between the AGN and the reflecting cloud, through
simple light crossing-time arguments: $\Delta l \sim c \Delta t_s$, where $\Delta l$ is
the difference in optical path between the primary and the reflected radiation. If
reflection occurs from the inner far side of the torus, $d \sim \Delta l/2$, where
$d$ is the distance between the reflecting cloud and the source of illuminating continuum.

If reflection occurs in an individual cloud
the difference in optical path length between photons coming from extreme regions of
the cloud has to be smaller than its light-crossing cloud linear size.
This puts a lower limit on its opening angle:
$$
\theta > arcsin( \frac{1}{1+S/d} )
$$
where $S$ is the size of the reflecting cloud and $d$ its distance from the primary source.
For $M_{pr}$: $S \le$0.05~pc, $d \approxgt$0.5~pc, $\theta \approxgt$71$^{\circ}$.

An independent estimate of the reflector opening
angle can be derived from the opening angle of the
ionization cone, $\theta_o$. Capetti et al. (1995) derive
$\theta_o >$55$^{\circ}$ from HST imaging of the NLR.
This estimate is larger then previous estimates based on ground-based
observations (\cite{wilson94}).

Ikeda et al. (2009) tried to constrain the geometry of the reflector by fitting the
{\it Suzaku}
time-average spectrum of Markarian~3 with Monte-Carlo simulations of the primary continuum
emission
by a Compton-thick accreting torus. The simultaneous measure of the reflected and of the
transmitted
(above 10~keV) components may in principle allow to constrain the geometry as well as the
column
density of the reflector (\cite{murphy09,yaqoob12}). Unfortunately, there is an intrinsic
degeneracy
between the opening angle and the inclination angle against the line-of-sight, which
cannot be resolved even with the best-quality broad band spectral data available. Awaki et al.
(2008) constrained the distance of the reflector within $\simeq$1~pc due to the low-ionization
and unresolved profile of the iron line. A similar analysis of the deepest EPIC-pn
spectrum by Bianchi et al. (2005) yielded an estimate of the inner size of the torus
comprised between 0.3 and 0.5~pc.

While the formal geometrical constrains on the torus derived by our variability study are not
inconsistent with those obtained by other studies using different techniques, one might ask
whether the implicit underlying picture of a single compact reflecting cloud with a size
smaller them 0.05~pc is tenable. We present in the following further observational evidence
which questions it.

\subsection{The K$_{\alpha}$ iron line emission region}

In the current
picture of Compton-thick reprocessing in the AGN environment, the Compton-reflection continuum
and the ``narrow''\footnote{We use the term ``narrow'' here to identify emission line features,
whose profile is not broadened and skewed by relativistic effects in X-ray illuminated accretion
disks a few gravitational radii from the black hole event horizon (\cite{fabian89})}
iron K$_{\alpha}$ fluorescent line should be produced by the same reflecting gas. Indeed, these
features almost invariably come together in the X-ray spectra of highly-obscured AGN. Existing
exceptions ({\it e.g.}, NGC7213, \cite{bianchi08}) are explained by reprocessing from
Compton-thin gas in BLR clouds. If both components are produced by
reprocessing of the same emission, one may expect them to respond simultaneously to changes
of the primary.

This does not seem to be the outcome of the X-ray spectroscopic campaign in Markarian~3. The intensity
of the strong and ubiquitous iron K$_{\alpha}$ fluorescent line does not follow the variability
of the Compton-reflected continuum (Fig.~\ref{fig2}). This finding is puzzling.
In order to explain it, one has to assume that the production of the line is decoupled
from the underlying continuum. The line could be
produced in Compton-thin clouds, while the Compton-reflection continuum could track Compton-thick
clouds. Several mechanisms can be invoked to explain the lack of a strong iron emission line
from Compton-thick material.
Ionization can suppress the emission line
via resonant trapping (\cite{matt93})
or full electron stripping.
However, the data with the best statistical quality rule out this explanation (cf. Fig.~\ref{fig12}).
Alternatively, the Compton-thick clouds could be iron deficient. It is, however, unclear why
abundances of different gaseous systems
in the environment of the same AGN should be largely different. Furthermore,
this hypothesis is inconsistent with the deep photoelectric absorption edge, clearly visible in
all spectra (see, {\it e.g.}, Fig.~\ref{fig7}). More importantly, there is no other evidence
of the presence of Compton-thin clouds from X-ray spectroscopic observations of Markarian~3,
although they could
not be detected along the same line-of-sight of the Compton-thick absorber.

\begin{figure}
\hspace{-0.5cm}
\includegraphics[height=90mm,angle=90]{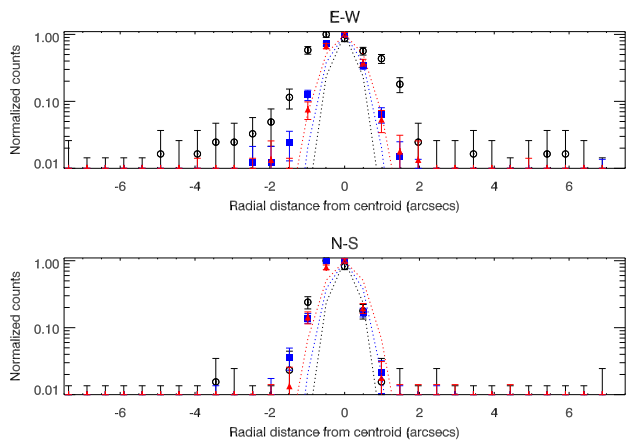}
\caption{
Intensity profiles measured from the 0$^{th}$ order image taken during the Chandra/HETG observation
of Markarian~3 (\cite{sako00}). From {\it top}: E-W; {\it bottom}: N-S,
Symbols indicate energy bands: {\it empty circles}: 0.5--3~keV; {\it
filled squares}: 3--6~keV; {\it filled triangles}: 6--7~keV. The {\it dotted lines}
indicate the Point Spread Function. From the {\it innermost} to the {\it outermost profiles} in each
panel: 0.5--3~keV, 3--6~keV, and 6--7~keV, respectively.
}
\label{fig8}
\end{figure}

To further investigate this point, we have reanalyzed the
0$^{th}$ order image of the
{\it Chandra}/HETG observation of Markarian~3 originally presented in Sako et al. (2000). As
they showed (cf. the Fig.1 in their paper), the soft X-ray emission is extended along an
approximate E-W direction (the elongation direction corresponds to an astronomical position
angle of 356$^{\circ}$.6). Intensity profiles in three different bands: 0.5--3~keV,
3--6~keV, and 6--7~keV are shown in Fig.~\ref{fig8}.
We confirm that emission is primarily  extended in the softest
energy band along the E-W direction up to $\simeq$4$\arcsec$ on the East side and $\simeq$2$\arcsec$
on the West side. In the hard band ($E \ge 3$~keV)
an excess by $\pm$1$\arcsec$ ($\simeq$270~pc)
is visible in the hardest bands along the
North direction only ({\it i.e.}, {\it perpendicular} to the extension of the 
X-ray NLR). However, there is no statistical difference between
the Compton continuum and the iron fluorescence line profiles.

Extended soft X-ray ($E \le 2$~keV)
emission on scales ranging from a few hundreds parsecs to a few thousands parsec has been
commonly discovered in highly obscured nearby AGN, almost invariably
in morphological agreement with the optical NLR (\cite{bianchi06} and references
therein). Extension in the hard energy band is much less common. Young et al. (2001) reported
hard X-ray extended emission, together with a tentatively detected iron line, in
NGC~1068 on scales $\simeq$20$\arcsec$ (2.2~kpc), alongside a low surface brightness component
up to $\simeq$50$\arcsec$ (5.5~kpc). More recently, Marinucci et al. (2012b) reported a clumpy
and flattened $\sim$150~pc structure in the hard band {\it Chandra}/ACIS image of NGC~4945.
Compact
hard X-ray emission in the Circinus~Galaxy (\cite{sambruna01}),
constrains the size of the Compton reflector to within 8~pc.
In bright nearby obscured AGN,
mir-IR interferometry allows to constraint the location and size of the dusty torus within
a few parsecs (\cite{jaffe04,meisenheimer07,tristram07}). While the X-ray and IR studies
indicate that dusty structures in the nuclear environment of AGN are extended on scales
from a fraction to hundred parsecs,
we stress that the two techniques probe two different phases of the obscuring matter in
the nuclear environment. X-ray direct imaging and variability are sensitive to the gaseous
phase, whereas IR measurements probe the dust component.

\subsection{Implications for the AGN structure model}

The results presented in this paper favour a complex structure of the optically-thick
absorber/reflector in Markarian~3. X-ray
energy-dependent cross-correlation results favor a reflecting
cloud at a distance $\approxgt$0.5~pc.
On the other hand, the variability of the Compton-reflection component on scales
$\approxlt$64--150~days (depending on the spectral model)
suggests that the variability episodes presented in this paper
might be dominated by small clouds,
implying a large opening angle, {\it i.e.} a small covering factor.
We discovered a puzzling lack correlation between the Compton-reflection flux and
the Fe K$_{\alpha}$ line intensity from our X-ray monitoring (Fig.~\ref{fig11};
correlation probability $\approxlt$10$^{-4}$). Furthermore,
at least part of the reflection occurs on scales as large as 300~pc (Fig.~\ref{fig8}).
Indirect evidence for optically thick reprocessing of the primary radiation with
a large covering fraction comes from the fact that
Markarian~3 is a strong mid-IR emitter. It is one of the prototypical AGN-dominated
Type~2 objects in mid-IR continuum (\cite{deo09}). This also represent an indirect evidence
in favour of an extended structure reprocessing the AGN emission into the IR band.

In summary, we report in this paper various (and partly conflicting) evidence
for optically-thick reprocessing of the AGN
primary emission in Markarian~3 on a variety of different spatial scales, from fraction
to hundreds of parsecs.

Comparing these findings with other Compton-thick AGN in the local Universe unveils a
similar variety, although 
detecting variability of the Compton-reflection component is difficult
and therefore rare.
In NGC~4945, the historical ({\it Chandra}, {\it Suzaku} and XMM-Newton)
light curve of the reflected component remains constant within $<$10\% despite
strong intrinsic variability of the Swift/BAT light curve on time scales longer than 1~year
(\cite{marinucci12b}).
Constant flux of the reflected components
over a monitoring campaign covering almost a decade
was reported in NGC~5506 (\cite{guainazzi10}.
On the other hand, a delayed response of the K$_{\alpha}$ iron line and Compton-reflection
to a decrease of the primary flux by a factor $\simeq$20 in 16~years constrained
the torus to be located at about 3.2~pc from the AGN in NGC~2992 (\cite{weaver96}).
Variability by about 25\% of the iron
K$_{\alpha}$ line on timescales of about 1~years were reported by different authors in
the unobscured Seyfert NGC~4151 (\cite{perola86,takahashi02,zdziarski02,schurch03}),
suggesting that the emitting region is at about 0.03~pc from the nucleus.

One possible explanation of these apparently contradicting results is that the absorber is clumpy.
Qualitatively, reflection on different spatial scales require that
absorption-free line-of-sights to the active nucleus must exist for clouds located at
various distances.
The case for a clumpy torus was originally advocated by
Krolik \& Begelman (1998), and more recently fully developed in the calculations by Nenkova et al.
(2002, 2008). While strong direct observational evidence exists for a clumpy absorber on scales of the
BLR in several nearby AGN
from X-ray variability studies (\cite{elvis04,risaliti05,puccetti07,bianchi09b,
maiolino10,risaliti11}), evidence for the torus clumpiness are mainly yielded by the comparison
between torus models and multiwavelength Spectral Energy Distributions. A compact, pc-scale
torus can hardly reproduce the multi-temperature IR emission (\cite{pier92}). Large-scale dusty
torii can reproduce the observations (\cite{granato94}), but
are nowadays disfavoured given the IR interferometric evidence of very compact warm dust structures
surrounding nearby obscured AGN (see, however, NGC~4945; \cite{marinucci12b}).

Recently, Eliztur (2012) proposed that the complex structure of the absorber/reflecting matter
in the AGN environment should lead us to a revision of our interpretation of the AGN Unification
Scenario. Its strictest formulation postulates that the orientation with respect to
an azimuthally-symmetric obscuring structure is the only driving factor in the classification
of an AGN. Elitzur suggests that
this view should be replaced by a {\it probabilistic} interpretation, whereby
the classification of a radio-quiet AGN would depend on the covering factor distribution
of the obscuring matter. ``Type~2'' would be more likely identified in AGN surrounded
by matter with a higher covering factor. A least extreme view, where the Poissonian 
fluctuations in the number of optically-thick clouds on scales comparable or lower then the BLR is
superposed to the standard formulation of the Unified Scenario
is discussed by Matt (2000) and Bianchi et al.
(2009).

Markarian~3 is one of the few obscured AGN in the local universe, where evidence of
AGN primary emission reprocessing
by optically thick gas on scales from a fraction of to several hundreds parsecs have been
reported. It may represent one of the best laboratories to test these scenarios with
future observations in X-rays as well as in other wavelengths. ALMA
is expected to give a paramount contribution to our understanding of the circumnuclear dusty torus.

\subsection{On the variability in the soft X-ray band}
\label{sect_soft}

The soft X-ray ($E \le 3$~keV) flux history in Markarian~3 also exhibit significant variations
around a mean of $\simeq$7$\times 10^{-13}$~erg~cm$^{-2}$~cm$^{-1}$.
The dynamical range is around $\pm$10\%.
This is of the same order of magnitude as the typical systematic uncertainties in the
flux determination between the EPIC-pn and the XRT detector (\cite{tsujimoto11}). Indeed,
the variability pattern is dominated by the comparison between the EPIC-pn and the XRT
observations, although the difference (XRT yielding systematically lower fluxes)
is {\it opposite} to what measured by \cite{tsujimoto11}.
A 5\% difference in the X-ray flux measured by XMM-Newton on 2000-2002 and 2012 could
be affected by uncertainties in the background subtraction in the most recent
observation (see Sect.~2). While we stress that further higher-quality measurements are 
required to confirm this finding, we discuss briefly in this Section some
of its possible implications.

Bianchi et al.
(2006) estimate that about 10\% of the emission in this band is unresolved in the {\it Chandra}/ACIS
image. A variability by a factor $\simeq$100\% of this compact component (associated to the
innermost region of the NLR) could account for the observed
variability pattern. Alternatively, the observed variability could be due to the spilling of the
primary emission through a partial covering absorber. In the Palermo BAT Catalogue (\cite{cusumano10})
Markarian~3 is detected with a 15--150~keV flux of
($1.18 \pm 0.03$)$\times 10^{-11}$~erg~cm$^{-2}$~cm$^{-1}$ and a spectral index $\Gamma_{BAT}=1.74$.
The extrapolation of this flux into the 0.5--1.5~keV energy band, assuming a simple power-law
of photon index $\Gamma_{BAT}$ is $\simeq$4.4$\times 10^{-11}$~erg~cm$^{-2}$~cm$^{-1}$.
The observed flux is about 1.5\% of this extrapolation.
Such a fraction is not unusual in Compton-thin AGN (\cite{risaliti02}).
See also Paggi et al. (2012) for the detection of variability in the soft
X-ray spectrum of Mkn~573.

\section{Conclusions}

The main conclusions of this paper can be summarised as follows:

\begin{itemize}

\item we report on a study of the decade-long
X-ray flux variability in Markarian~3 in the 4--10~keV
energy band, dominated by Compton-reflection of the primary AGN emission. The shortest variability
timescale is $\approxlt$64~days, as measured by the shortest interval between two statistically
inconsistent measurements

\item assuming that the light curve of the Compton-reflection dominated
energy band and that of the 15--150~keV energy band are correlated, with the former lagging the
latter, the minimum measured delay is $\approxgt$1200~days

\item the intensity of the iron K$_{\alpha}$ emission line
is uncorrelated to the flux in the Compton-reflection dominated band. This suggest that
the bulk of the iron line is emitted in a different, more extended structure. Energy-resolved
analysis of existing {\it Chandra} data indicates a region of extended emission of both
the line-free Compton-reflection continuum and of the Fe K$_{\alpha}$ line up to
$\sim$300~pc to the North of the active nucleus

\end{itemize}

Markarian~3 exhibit a rich phenomenology associated to the optically-thick reprocessing of
the AGN primary radiation. Evidence of reprocessing on very different spatial scales have
been presented and discussed in this paper. They flow with the stream of mounting experimental
evidence in favor of an extended and clumpy structure of the absorbing/reprocessing gas and
dust in the innermost parsecs around the super-massive black hole. Revisions of the standard
interpretation of the Unified Scenario encompassing this complexity have been suggested
by several authors (\cite{matt00,elvis00,elitzur12}).

Still, the constraints on the geometry of the reprocessing matter in Markarian~3 are
dependent on the details of the spectral modeling, due to
the coarse X-ray monitoring in the Compton-reflection dominated energy band, as well as to
the statistical quality of the available data. We intend
to tighten them in the nearby future, by pursuing a specific observational monitoring program.
The perspectives offered by the successful launch of NuSTAR (\cite{harrison10})
are particularly promising.
It is reasonable to assume that at least part of the material reflecting the intrinsic continuum
is also responsible for the absorption along the line-of-sight. In this case variability of the
measured line-of-sight may provide additional constraints. These measurement are
virtually impossible now, but may be possible thanks to the increased sensitivity
of NuSTAR at energies larger then the photoelectric cut-off. Similar measurements may be possible
also with Astro-H in a nearby future.

\section{Appendix~A}

\begin{figure*}
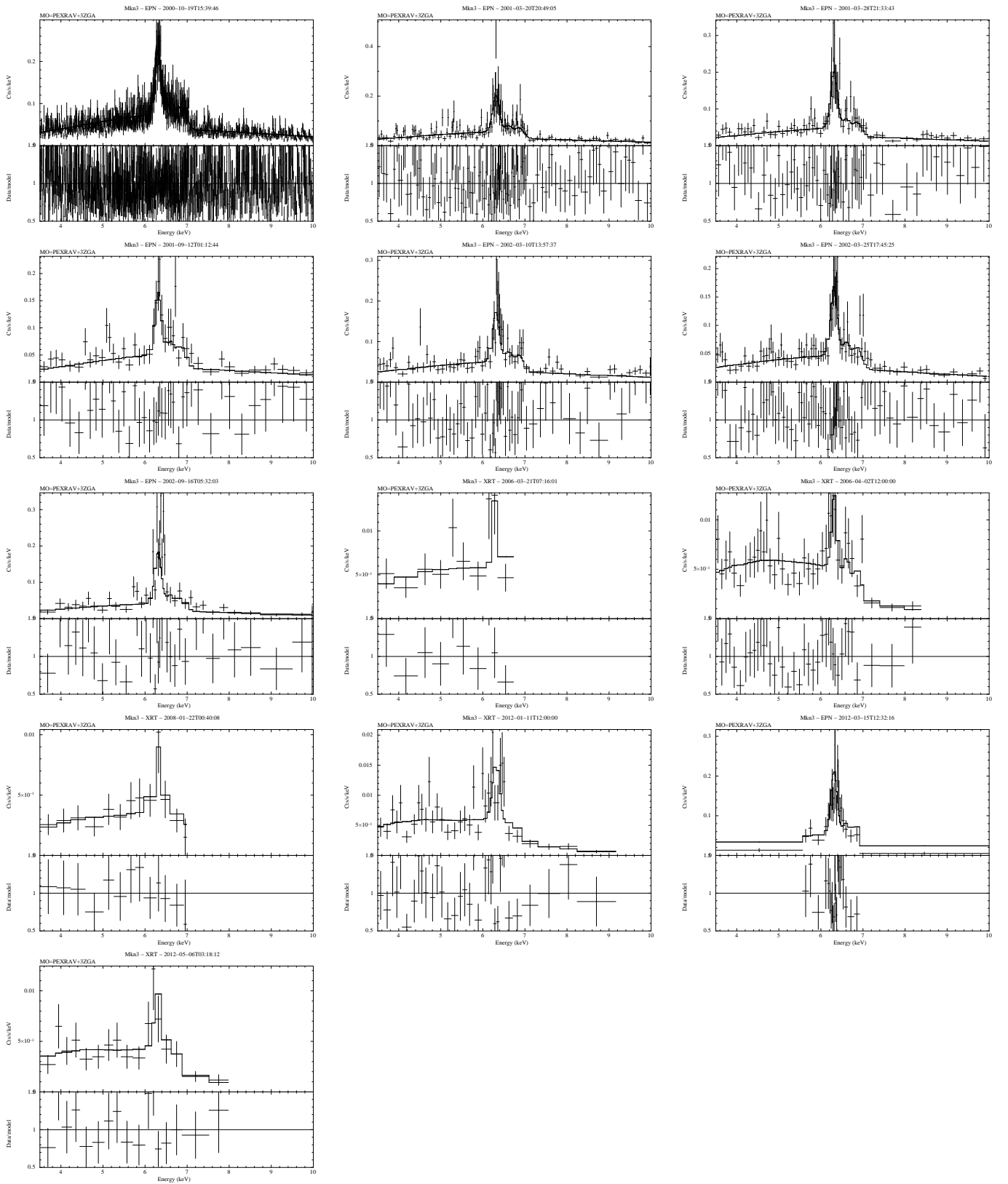

\hbox{
\includegraphics[height=60mm,angle=-90]{0111220201_pn_p04_r40_gti04_best_fit.ps}
\includegraphics[height=60mm,angle=-90]{0009220601_pn_p04_r40_gti04_best_fit.ps}
\includegraphics[height=60mm,angle=-90]{0009220701_pn_p04_r40_gti04_best_fit.ps}
}
\hbox{
\includegraphics[height=60mm,angle=-90]{0009220901_pn_p04_r40_gti04_best_fit.ps}
\includegraphics[height=60mm,angle=-90]{0009220401_pn_p04_r40_gti04_best_fit.ps}
\includegraphics[height=60mm,angle=-90]{0009220501_pn_p04_r40_gti04_best_fit.ps}
}
\hbox{
\includegraphics[height=60mm,angle=-90]{0009221601_pn_p04_r40_gti04_best_fit.ps}
\includegraphics[height=60mm,angle=-90]{2006_03_21_best_fit.ps}
\includegraphics[height=60mm,angle=-90]{2006_04_best_fit.ps}
}
\hbox{
\includegraphics[height=60mm,angle=-90]{2008_01_22_best_fit.ps}
\includegraphics[height=60mm,angle=-90]{2012_01_best_fit.ps}
\includegraphics[height=60mm,angle=-90]{0656580301_pn_src_best_fit.ps}
}
\hbox{
\includegraphics[height=60mm,angle=-90]{2012_05_06_best_fit.ps}
}
\caption{
Spectra ({\it upper panels}) and residuals in units of data/model ratio ({\it lower panels}) when
the baseline model is applied to the spectra of the observations discussed in this paper.
The data were rebinned such that each data
point corresponds to a sigma-to-noise ratio $>$3 for visualization purposes only (the
analysis was performed on the unbinned spectra). The spectrum of the deepest observation (October 2000;
{\it upper left}) is shown with a coarser binning (minimum signal-to-noise ratio $>$10) in
Fig.~\ref{fig12}.
}
\label{fig7}
\end{figure*}

In Fig.~\ref{fig7} we show the spectra ({\it upper panels}) and residuals in units of data/model ratio
({\it lower panels}) for all the observations analyzed in this paper.

\begin{acknowledgements}
Based on observations obtained with XMM-Newton, an ESA science mission with instruments and
contributions directly funded by ESA Member States and NASA. Financial support for this work was
partly provided by the Spanish Ministry of Science and Innovation through grant AYA2010-21490-C02-02.
We are grateful to the referee, whose careful revision of the manuscript lead to a substantial
change in the structure of the paper, as well as to a more careful assessment of the astrophysical
implication of our results.
\end{acknowledgements}


\begin{thebibliography}{} 

\bibitem[Alonso-Herrero et al., 2011]{alonsoherrero11} Alonso-Herrero A., Ramos-Almeida C., Mason R., et al., 2011, ApJ, 736, 82

\bibitem[Anders \& Grevesse 1989]{anders89} Anders E. \& Grevesse N., 1989, Geochimica et Cosmochimica Acta 53, 197

\bibitem[Antonucci 1993]{antonucci93} Antonucci R., 1993, ARA\&A 31, 473

\bibitem[Antonucci \& Miller 1985]{antonucci85} Antonucci R.R.J., Miller J.S., 1985, ApJ 297, 621

\bibitem[Awaki et al. 1991]{awaki91} Awaki H., Koyama K., Inoue H., Halpern J.O., 1991, PASJ 43, 195

\bibitem[Awaki et al. 2008]{awaki08} Awaki H., Anabuki N., Fukazawa Y., et al., 2008, PASJ, 60, S293

\bibitem[Barr \& Mushotzky 1986]{barr86} Barr P., Mushotzky R.F., 1986, Nat, 320, 421

\bibitem[Barthelmy et al. 2005]{barthelmy05} Barthelmy S.~D.,
Barbier L.~M., Cummings J.~R., et al., 2005, SSR, 120, 143

\bibitem[Barvainis 1987]{barvainis87} Barvainis R., 1987, ApJ, 320, 537

\bibitem[Bennett et al. 2003]{bennett03} Bennett C.L., et al., 2003, ApJS, 148, 1

\bibitem[Bianchi et al. 2006]{bianchi06} Bianchi S., Guainazzi M., Chiaberge M., 2006, A\&A, 448, 499

\bibitem[Bianchi et al. 2008]{bianchi08} Bianchi S., La Franca F., Matt G., et al., 2008, MNRAS, 389, L52

\bibitem[Bianchi et al. 2005]{bianchi05} Bianchi S., Miniutti G., Fabian A.C., Iwasawa K., 2005, MNRAS, 360, 380

\bibitem[Bianchi et al. 2009a]{bianchi09a} Bianchi S., Guainazzi M., Matt G., Fonseca Bonilla N., Ponti G., A\&A, 2009, 495, 421

\bibitem[Bianchi et al. 2009b]{bianchi09b} Bianchi S., Piconcelli E., Chiaberge M., Jim\'enez-Bail\'on E., Matt G., Fiore F., 2009, ApJ, 695, 781

\bibitem[Bianchi et al. 2012]{bianchi12} Bianchi S., Maiolino R., Risaliti G., 2012, AdAST, 2012, 17

\bibitem[Burrows et al. 2005]{burrows05} Burrows D.~N., Hill J.~E., Nousek J.~A., et al., 2005, SSR, 120, 165

\bibitem[Burtscher et al. 2009]{burtscher09} Burtscher L., Jaffe W., R\"ottgering H., Meisenheimer K., Tristram K.R.W., 2009, MNRAS, 394, 1325

\bibitem[Capetti et al., 1995]{capetti95} Capetti A., Macchetto F., Axon D.J., Sparks W.B., Boksenberg A., 1995, ApJ, 448, 600

\bibitem[Cappi et al. 1999]{cappi99} Cappi M., et al., 1999, A\&A, 344, 857

\bibitem[Cash 1976]{cash76} Cash W., 1976, A\&A, 52, 307

\bibitem[Cusumano et al. 2010]{cusumano10} Cusumano G., La Parola V., Segreto A., et al., 2010, A\&A, 510, 48

\bibitem[Dadina 2007]{dadina07} Dadina M., 2007, A\&A, 461, 1209

\bibitem[Deo et al. 2009]{deo09} Deo R.P., Richards T.G., Crenshaw D.M., Kraemer S.B., 2009, ApJ, 705, 14

\bibitem[Elitzur 2012]{elitzur12} Elitzur M., 2012, ApJ, 747, L33

\bibitem[Elvis 2000]{elvis00} Elvis M., 2000, ApJ, 545, 63

\bibitem[Elvis et al. 2004]{elvis04} Elvis M., Risaliti G., Nicastro F., Miller J.M., Fiore F., Piccetti S., 2004, ApJ, 615, L25

\bibitem[Fabian et al. 1989]{fabian89} Fabian A.C., Rees M.J., Stella
L., White N.E., 1989, MNRAS, 238, 729

\bibitem[Gabriel et al. 2003]{gabriel03} Gabriel C., Denby M., Fyfe D. J., Hoar J., Ibarra A., 2003, in ASP Conf. Ser., Vol. 314 Astronomical Data Analysis Software and Systems XIII, eds. F. Ochsenbein, M. Allen, \& D. Egret (San Francisco: ASP), 759 

\bibitem[Gehrels et al. 2004]{gehrels04} Gehrels N., Chincarini G., Giommi P., et al., 2004,
ApJ, 611, 1005

\bibitem[Georgantopoulos et al. 1999]{georgantopoulos99} Georgantopoulos I., Papadakis I., Warwick R.S., Smith D.A., Stewart G.C, Griffiths R.G, 1999, MNRAS, 307, 815

\bibitem[Granato \& Danese 1994]{granato94} Granato G.L., Danese L., 1994, MNRAS, 268, 235

\bibitem[Guainazzi et al. 2010]{guainazzi10} Guainazzi M., Bianchi S., Matt G., et al., 2010, MNRAS, 406, 2013

\bibitem[Harrison et al. 2010]{harrison10} Harrison F.A., et al. 2010, SPIE, 7732, 27

\bibitem[Krolik \& Begelman 1988]{krolik88} Krolik J.H., Begelman M.C., 1988, ApJ, 329, 702

\bibitem[Jaffe et al. 2004]{jaffe04} Jaffe W., Meisenheimer K., R\"ottgering H.J.A., et al., 2004, Nat, 429, 47

\bibitem[Ikeda et al. 2009]{ikeda09} Ikeda S., Awaki H., Terashima Y, 2009, ApJ, 692, 608

\bibitem[Iwasawa et al. 1994]{iwasawa94} Iwasawa K., Yaqoob T., Awaki H., Ogasaka Y., 1994, PASJ, 46, L167

\bibitem[Lamastra et al. 2008]{lamastra08} Lamastra A., Perola G.C., Matt G., 2008, A\&A, 487, 109

\bibitem[Magdziarz \& Zdziarski 1995]{magdziarz95} Magdziarz P. \& Zdziarski A.A., 1995, MNRAS 273, 837

\bibitem[Maiolino et al. 2010]{maiolino10} Maiolino R., Risaliti G., Salvati M., et al., 2010, A\&A, 517, 47

\bibitem[Marinucci et al. 2012a]{marinucci12a} Marinucci A., Bianchi S., Nicastro F., Matt G., Goulding A.D., 2012, ApJ, 748, 130

\bibitem[Marinucci et al. 2012b]{marinucci12b} Marinucci A., Risaliti G., Wang Junfeng, et al., 2012, MNRAS, 423, L6

\bibitem[Mateos et al. 2010]{mateos10} Mateos S., Carrera F.J., Page M.J., et al., 2010, A\&A, 510, 35

\bibitem[Matt 2000]{matt00} Matt G., 2000, A\&A, 335, L31

\bibitem[Matt et al. 1993]{matt93} Matt G., Fabian A.C., Ross R.R., 1993, MNRAS 261, 346

\bibitem[Matt et al. 2003]{matt03} Matt G., Guainazzi M., Maiolino R., 2003, MNRAS, 342, 422

\bibitem[McHardy et al. 2006]{mchardy06} McHardy I., Koerding E., Knigge C., Uttley P., Fender R.P., 2006, Nat, 444, 730

\bibitem[Meisenheimer et al. 2007]{meisenheimer07} Meisenheimer K., Tristram K.R.W., Jaffe W., et al., 2007, A\&A, 471, 453

\bibitem[Murphy \& Yaqoob 2009]{murphy09} Murphy K., Yaqoob T., 2009, MNRAS, 397, 1549

\bibitem[Nenkova et al. 2002]{nenkova02} Nenkova M., Ivezi\'c Z., Elitzur M., 2002, ApJ, 570, L9

\bibitem[Nenkova et al. 2008]{nenkova08} Nenkova M., Sirocky M.M., Nikutta R., Ivezi\'c Z., Elitzur M., 2008, 685, 160

\bibitem[Nicastro 2000]{nicastro00} Nicastro F., 2000, ApJ, 530, L65

\bibitem[Paggi et al., 2012]{paggi12} Paggi A., Wang Junfeng, Fabbiano G., Elvis M., Karvoska M., 2012, ApJ, in press (arXiv:1203.1279)

\bibitem[Perola et al. 1986]{perola86} Perola G.C., Piro L., Altamore A., et al., 1986, ApJ, 306, 508

\bibitem[Peterson et al. 2004]{peterson04} Peterson B.M., Ferrarese L., Gilbert K.M., et al., 2004, ApJ, 613, 682

\bibitem[Pier \& Krolik 1992]{pier92} Pier E.A., Krolik J.H., 1992, ApJ 399, L23

\bibitem[Pounds et al. 2005]{pounds05} Pounds K.A., Page K.L., 2005, MNRAS, 360, 1123

\bibitem[Puccetti et al. 2007]{puccetti07} Puccetti S., Fiore F., d'Elia V., et al., 2007, MNRAS, 377, 607

\bibitem[Risaliti 2002]{risaliti02} Risaliti G., 2002, A\&A, 386, 379

\bibitem[Risaliti et al. 2005]{risaliti05} Risaliti G., Elvis M., Fabbiano G., Baldi A., Zezas A., 2005, ApJ, 623, L93

\bibitem[Risaliti et al. 2011]{risaliti11} Risaliti G., Nardini E., Salvati M., et al., 2011, MNRAS, 410, 1027

\bibitem[Ross \& Fabian 2005]{ross05} Ross R.R., Fabian A.C., 2005, MNRAS, 358, 211

\bibitem[Sako et al. 2000]{sako00} Sako M., Kahn S.M., Paerels F., Liedahl D.A., 2000, ApJL 543, L115

\bibitem[Sambruna et al. 2001]{sambruna01} Sambruna R., Netzer H., Kaspi S., et al., 2001, ApJ, 546, L13

\bibitem[Sartore et al. 2012]{sartore12} Sartore N., Tiengo A., Mereghetti S., De Luca A., Turolla R., Haberl F., 2012, A\&A, 541, 66

\bibitem[Schurch et al. 2003]{schurch03} Schurch N.J., Warwick R.S., Griffiths R.E., Sembay S., 2003, MNRAS, 345, 423

\bibitem[Segreto et al. 2010]{segreto10} Segreto A., Cusumano G., Ferrigno C., et al., 2010, A\&A, 510, 47

\bibitem[Str\"uder et al. 2001]{struder01} Str\"uder L., Briel U., Dennerl K., et al., 2001, A\&A 365, L18

\bibitem[Takahashi et al. 2002]{takahashi02} Takahashi K., Inoue H., Dotani T., 2002, PASJ, 54, 373

\bibitem[Tristram et al., 2007]{tristram07} Tristram K.R.W., Meisenheimer K., Jaffe W., et al., 2007, A\&A, 474, 837

\bibitem[Tsujimoto et al. 2011]{tsujimoto11} Tsujimoto M., Guainazzi M., Plucinsky P.P., et al., 2011, A\&A, 525, 25

\bibitem[Yaqoob 2012]{yaqoob12} Yaqoob T., 2012, MNRAS, in press (astro-ph/1204.4196)

\bibitem[Young et al. 2001]{young01} Young A.J., Wilson A.S., Shopbell P.L., 2001, ApJ, 556, 6

\bibitem[Weaver et al. 1996]{weaver96} Weave, K.A., Nousek, J., Yaqoob T., Mushotzky R.~F., Makino F., Otani C., 1996, ApJ, 458, 160

\bibitem[Wilms et al. 2000]{wilms00} Wilms J., Allen A., McCray R., 2000, ApJ, 542, 914

\bibitem[Wilson \& Tsvetanov 1994]{wilson94} Wilson A.S., Tsvetanov Z.I., 1994, AJ, 107, 1227

\bibitem[Wilson \& Ulvestad 1983]{wilson83} Wilson A.S., Ulvestad J.S., 1983, ApJ, 275, 8

\bibitem[Zdziarski et al. 2002]{zdziarski02} Zdziarski A.A., Leighly K.M., Matsuoka M., Cappi M., Mihara T., 2002, ApJ, 573, 505

\end{thebibliography}
\end{document}